\documentstyle[aaspp]{article}

\input epsf.sty
\tightenlines

\begin{document}

\slugcomment{Submitted to the Astrophysical Journal, February 1998}

\title{\LARGE\bf Three types of gamma ray bursts}

\vspace{0.3in}

\author{\Large Soma Mukherjee$^{1,2,3}$, Eric D. Feigelson$^4$, Gutti Jogesh
Babu$^5$, Fionn Murtagh$^{6,7}$, Chris Fraley$^8$ and Adrian
Raftery$^8$}  
\affil{1. Indian Statistical Institute, 203 Barrackpore Trunk Road, Calcutta         
  700 035, India}
\affil{2. Department of Physics \& Astronomy, Northwestern University,
Evanston IL 60208-3112}
\affil{3. Present address: LIGO Project, California Institute of  Technology,
MS 18-34, Pasadena CA 91125}
\affil{4. Department of Astronomy \& Astrophysics, 525 Davey Laboratory,  
Pennsylvania State University, University Park PA 16802. Email:
edf@astro.psu.edu.   To whom correspondance should be addressed.} 
\affil{5.  Department of Statistics, 326 Thomas Building, Pennsylvania State   
University, University Park PA 16802.}
\affil{6. Faculty of Informatics, University of Ulster, North Ireland, United
Kingdom.} 
\affil{7. Observatoire Astronomique, 11 rue de l'Universit\'e, F-67000
Strasbourg, France}
\affil{8. Department of Statistics, University of Washington, Box 354322,    
Seattle WA 98195-4322} 

\vspace{0.3in}

\begin{abstract}

A multivariate analysis of gamma-ray burst (GRB) bulk properties is presented
to discriminate between distinct classes of GRBs.  Several variables representing  burst
duration, fluence and spectral hardness are considered.  Two multivariate
clustering procedures are used on a sample of 797 bursts from the Third
BATSE Catalog: a nonparametric average linkage hierarchical agglomerative
clustering procedure validated with Wilks' $\Lambda^*$ and other MANOVA
tests; and a parametric maximum likelihood model-based clustering procedure
assuming multinormal populations calculated with the EM Algorithm and
validated with the Bayesian Information Criterion.  

The two methods yield very similar results.  The BATSE GRB
population consists of three classes with the following
Duration/Fluence/Spectrum bulk properties: Class I with 
long/bright/intermediate bursts, Class II with short/hard/faint bursts, and Class
III with intermediate/intermediate/soft bursts.  One outlier with poor data is also
present.  Classes I and II  correspond to those reported by Kouveliotou et al.
(1993), but Class III  is clearly defined here for the first time. 
\end{abstract}

\keywords{gamma rays: bursts; methods: statistical; methods: data analysis}  

\clearpage

\section{Introduction}

As very few gamma-ray burst (GRB) sources have astronomical counterparts at
other  wavebands, empirical studies of GRBs have been largely restricted to the
analysis of their gamma ray properties: bulk properties such as fluence and
spectral hardness, and evolution of these properties within a burst event
(Fishman \& Meegan 1995).  While bursts exhibit a vast range of complex
temporal behaviors, their bulk properties appear simpler and amenable to
straightforward  statistical analyses.  Studies fall into two categories:
examination of whether GRB bulk properties comprise a homogeneous population
or are divided into distinct classes; and search for relationships between bulk
properties.  Both types of study may lead to astrophysical insight, just as the
distinction between main sequence stars and red giants and the measurement of
a luminosity-mass relation along the main sequence assisted the development of
stellar astrophysics early in the century.

The most widely accepted taxonomy of GRBs is the division between 
short-hard and long-soft bursts proposed by Dezalay et al. (1992) and
Kouveliotou et al. (1993, henceforth K93).  K93 noticed a bimodality in the
burst duration variable $T_{90}$ (time within which 90\% of the flux arrived),
suggesting the presence of two distinct types of bursts separated at $T_{90}
\simeq 2$ sec.  The short bursts have systematically harder gamma-ray spectra
than longer bursts. The two groups seemed indistinguishable in most other bulk
properties, although the larger group of long-soft bursts may have a
subclass with a different fluence distribution (i.e., different $<V/V_m>$; Katz
\& Canel 1996) and the groups may have different Galactic latitude
distributions (Belli 1997).  Other researchers point to small groups of bursts
with distinctive properties such as the soft-gamma repeaters (Norris et al.
1991),  two possible classes with differing short-timescale variability (Lamb,
Graziani  \& Smith 1993), fast-rise exponential-decay bursts (Bhat et al. 1994),
and two  types of bursts with different ratios of total fluence and $>$300 keV
fluence  (Pendleton et al. 1997).

A variety of relationships between burst properties have also been reported. 
Norris et al. (1995) find an anti-correlation between $T_{90}$ (calculated after
wavelet thresholding) and peak intensity, consistent with a cosmological time
dilation. However, a positive correlation between $T_{90}$ and total fluence is
also seen which does not agree with the simplest cosmological interpretation
(Lee \& Petrosian 1997).  Additional reported relationships include: $T_{90}$ 
correlated with peak heights (Lestrade 1994), peak energy correlated with peak 
flux (Mallozzi et al. 1995), and peak duration anticorrelated with gamma-ray
energy (Fenimore et al. 1995).

Most of these studies suffer from a failure to treat all of the bulk property
variables in an unbiased and quantitative way.  Astronomers typically examine
univariate or bivariate distributions, sometimes constructing composite variables
(such as hardness ratios) with pre-determined relationships to include one or
two additional variables.  But it is quite possible that the complex astrophysics
producing GRBs will not manifest themselves in simple bivariate plots, just as
the division between short-hard and long-soft bursts is not evident in spectral
variables alone (Pendleton et al. 1994). GRB catalogs, like most
multiwavelength astronomical catalogs, are multivariate databases and should be
treated with multivariate statistical methods that can objectively and effectively
uncover structure involving many variables (Feigelson \& Babu 1997).    Two
previous studies take a fully multivariate approach to understanding GRB bulk
properties.  Baumgart (1994) constructs a neural network taxonomy of 99
GRBs from the PVO satellite using 26 variables representing both bulk burst
properties and detailed temporal characteristics (e.g. number of peaks, fractal
dimension, wavelet transform crossings) and finds two or three distinct GRB
classes. Bagoly et al. (1997) perform principal components and factor analyses
of nine bulk property variables using 625 GRBs from the BATSE 3B catalog. 
They find that the relationships in the database are determined principally by
only three variables: an appropriately weighted fluence, a weighted burst
duration, and (to a lesser extent) flux in the highest energy bin.

We note, however, that it can be dangerous to look for correlations prior to 
classifying (or establishing the homogeneity of) the population.  While the 
anticorrelation between hardness ratio and burst duration seen in full samples 
(K93) may be the manifestation of a single astrophysical process, it may 
alternatively reflect differences between distinct processes.  The latter 
possibility is suggested by a reported hardness-duration positive correlation
within  the long-soft class of bursts (Dezalay et al.  1996; Horack \& Hakkila
1997).   Most multivariate analyses thus begin with a study of homogeneity and 
classification, and then investigate the variance-covariance structure (i.e. 
correlations) within each class.    

This paper describes a multivariate analysis of GRBs from the Third BATSE
Catalog (Meegan et al. 1996).  After defining the sample (\S 2), we start with a
simple statistical description of the variables and their bivariate relationships for
the entire dataset (\S 3).  We then seek distinct types of clusters in two ways. 
First, a standard nonparametric agglomerative hierarchical clustering analysis is
performed (\S 4) which reveals three distinct classes.  The statistical
significance of  the third cluster is validated, under Gaussian assumptions, with
MANOVA tests.   Second, a parametric maximum likelihood model-based
clustering procedure is adopted which reveals the same three groups and
indicates strong evidence for the presence for three rather than two groups (\S
5).  The variance-covariance structure of each group is then examined (\S 6).
Results are synthesized in the discussion (\S 7).   

Throughout the paper, we discuss our mathematical techniques to help the 
reader understand the complexities of  multivariate analysis.  From the vast 
literature in this subject, we recommend the following monographs for
interested  readers:  Johnson \& Wichern (1992) and Jobson (1992) for
overviews of applied  multivariate analysis;  Hartigan (1975), Jain \& Dubes
(1988) and Kaufman \&  Rousseeuw (1990) for multivariate clustering
algorithms;   Murtagh \& Heck  (1987) and, more briefly,  Babu \& Feigelson
(1996) and Feigelson \& Babu (1997), for multivariate methodology  in
astronomy. 

\section{The GRB sample and statistical software}

Our sample is drawn from the Third Catalog of the {\it Burst and Transient
Source Experiment} (BATSE) on board the Compton Gamma Ray
Observatory.   This 3B catalog has 1122 GRBs detected by BATSE between
1991 April 19 and 1994 September 19.  The catalog is presented and fully
described by Meegan et al. (1996).  Our database was extracted from the
on-line database www.batse.msfc.nasa.gov/data/grb/catalog in May 1996,
which provides many properties of each burst.  There are roughly eleven
variables of potential astrophysical interest: two measures of location in Galactic
coordinates, $l$ and $b$; two measures of burst durations, the times within
which 50\% ($T_{50}$) and 90\% ($T_{90}$) of the flux arrives; three peak
fluxes $P_{64}$, $P_{256}$ and $P_{1024}$ measured in 64 ms, 256 ms and
1024 ms bins respectively; and four time-integrated fluences $F_1$-$F_4$ in
the 20-50 keV, 50-100 keV, 100-300 keV and $>300$ keV spectral channels
respectively.  Researchers commonly consider three composite variables: the
total fluence, $F_T = F_1+F_2+F_3+F_4$,  and two measures of spectral
hardness derived from the ratios of channel fluences, $H_{32} = F_3/F_2$ and
$H_{321}= F_3/(F_1+F_2)$.  Due to the  limitations of available multivariate
statistical techniques, we ignore other variables of potential relevance including
the heteroscedastic measurement errors of each quantity (i.e., errors that differ
from point to point) and truncation values associated with BATSE triggering
operations. 

Of the 1122 listed bursts, 807 have data on all the variables described above. 
Ten bursts listed with zero fluences were eliminated.  Our sample thus has 797
BATSE GRBs.   For some analyses, we also used a subset of 644 bursts with
`debiased' durations, $T^d_{90}$.  Here the durations are modified to account
for the effect that brighter bursts will have signal above the noise for longer
periods than fainter bursts with the same time profiles (J. Norris, private
communication). 

Statistical analyses in \S\S 3, 4 and 6 were conducted within the Statistical
Analysis System {\it SAS/STAT\footnote{{\it SAS/STAT} is a registered
trademark of the SAS Institute Inc.}}, a very large and widely distributed commercial
statistical software package (SAS Institute Inc. 1989).  {\it SAS/STAT}
procedures {\it CLUSTER},  {\it GLM}, {\it PRINCOMP} and {\it
VARCLUS} were used.  The analysis in \S 5 was performed with the {\it
MCLUST} software (Banfield \& Raftery 1993; Fraley  1998), which is
interfaced to the Splus statistical package (Splus Version 3.4;
MathSoft, Inc.\ 1996) and its extensions. Further information is provided at
http://stat.washington.edu/fraley/software.html
and http://lib.stat.cmu.edu/general/mclust.  For multivariate data
visualization, we used the  XGobi (Swayne, Cook \& Buja 1991) program, 
available from the {\it StatLib} software archive
at http://lib.stat.cmu.edu/general/XGobi. Hypertext
links to a variety of public domain software for multivariate analysis,
classification and visualization are available at the Penn State {\it StatCodes}
Web site, http://www.astro.psu.edu/statcodes.  

\section{Statistical properties of the entire sample}

We are faced with a multivariate database of 797 objects and 15 variables (11
variables from the catalog, 3 composite variables, and $T_{90}^d$). Two
initial problems are frequently faced in analyses of multivariate databases. First,
variables with incompatible units and ranges must be compared. Units can be
removed by normalization (e.g. replacing $F_1$ by $F_1/F_{tot}$), by 
standardization (e.g. replacing $F_1$ by $F_1/\sigma$ where $\sigma$ is the
sample standard deviation), or by  taking logarithms.  Second, the dimensionality 
of the problem should be reduced, as many of the variables are closely
interrelated either by construction or by astronomical circumstance.  Although
there are  no mathematical rules regulating reduction of dimensionality, it can
usefully be guided by a correlation matrix showing bivariate relationships and a
principal components analysis showing multivariate relationships that are mainly
responsible for structured variance in the data.  Scientific reasoning can also be
used to eliminate consideration of variables.   We conducted a preliminary
examination of data representations, correlation  matrices and bivariate plots,
and principal components analyses to facilitate choice of variables.  When no
mathematical preference arose, we selected variables  most commonly used by
previous researchers to facilitate comparison of results. 

Our choices were as follows.  We use log variables, rather than  normalized or
standardized variables.  We kept information on burst fluence and spectra
through $F_{tot}$ and hardness ratios rather than through the original fluences
$F_1$-$F_4$.  We initially eliminated $P_{64}$ and $P_{1024}$ from
consideration, and later eliminated $P_{256}$ when we found that its main
contribution to the clustering process was noise.  We chose to remove the location
variables $(l,b)$, already established by other researchers to be random for the
entire sample, but use them later to test for isotropy of subsamples.  The
debiased $T_{90,d}$  is used only in special tests.  Our analysis was thus
performed in six or fewer dimensions using log $T_{50}$, log $T_{90}$, log
$F_{tot}$, log $P_{256}$, log $H_{321}$ and log $H_{32}$.   

\begin{figure}
\plotone{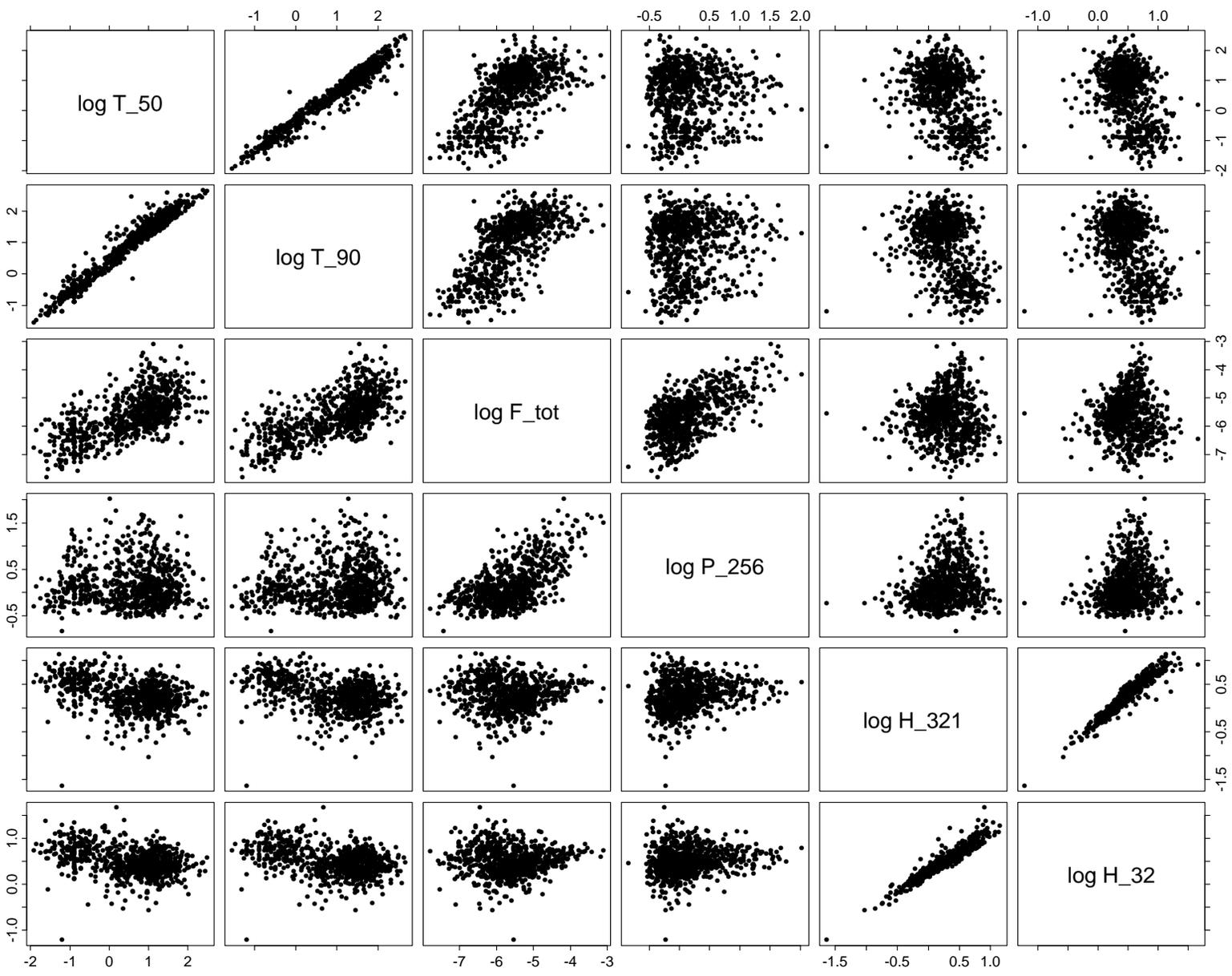}
\caption{Mosaic of scatter plots of six bulk properties for the 797
GRBs from the Third BATSE Catalog used in this study.}
\end{figure}

Tables 1 and 2 give basic statistics for these six variables:  means, standard
deviations, and bivariate values of Pearson's linear correlation  coefficient $r$. 
For $N = 797$ and assuming  bivariate normal populations, any $|r| > 0.013$
implies that a correlation between the  two variables exists at a two-tailed
significance level $P < 0.001$ (Beyer 1968,  pp. 389 and 283).  But from an
astrophysical perspective, we might consider any relationship with $|r| \lesssim
0.1$ to be of little interest.  Figure 1 shows the  bivariate scatter plots.

The correlation structure of the entire sample (Table 2) shows that the two
measures  of duration and the two measures of spectral hardness have
correlation near unity, indicating that they are nearly redundant.  $F_{tot}$ and
$P_{256}$ are quite dissimilar: $F_{tot}$ shows a strong correlation with burst
duration (e.g. Lee \& Petrosian 1997)  and no relation to hardness, while
$P_{256}$ shows no relation to duration but is mildly  correlated with hardness
(Mallozzi et al. 1995). Burst duration is anticorrelated with  hardness (K93;
Fenimore et al. 1995).  The cosmological anticorrelation between duration  and
peak flux reported by Norris et al. (1995) is statistically significant, but accounts 
for only a few percent of the variance between these variables.   The correlation
matrix based on the debiased $T_{90}^d$ values yields very similar results. 

However, the scatter plots (Figure 1) show a more complex story.  First, many plots show
inhomogeneous distributions inconsistent with the unimodal multinormal  (i.e.
multivariate Gaussian) population assumed by Pearson's $r$. The distributions
often seem bimodal with asymmetrical non-Gaussian shapes.  One outlier burst
is also seen in several projections.   We therefore consider the hypothesis that
the sample consists of two or more distinct classes, and proceed to find the
`clusters' using well-established methods.

\section{Nonparametric hierarchical cluster analysis}

\subsection{Methodological background}

Agglomerative hierarchical clustering is a procedure based on the successive
merging of proximate pairs of clusters of objects.  It produces a clustering tree
or dendrogram starting with N clusters  of 1 member (or a coarse partition
based on prior knowledge) and ending with one cluster of N members.
Unfortunately, there are many possible ways to proceed; mathematics provides
little guidance among the choices and no probabilistic evaluation of the results
without the imposition of additional assumptions.   The scientist must make
four decisions to fully define the clustering procedure: 
\begin{enumerate}

\item Creating {\it unit-free variables} is essential for meaningful treatment of
objects in multivariate space (\S 3).   A favorite choice by statisticians is
standardization, where each variable is normalized by the standard deviation of
the sample.  Astronomers more commonly make logarithmic transformations or
construct ratios of variables sharing the same units.   We follow the tradition of
GRB researchers by measuring spectral hardness with ratios of fluences having
the same units, and making  logarithmic transformations of all variables. 

\item The {\it metric} defines the meaning of proximity between two objects or
clusters.  Common choices are the simple Euclidean distance between unit-free
variables and the squares of Euclidean distances.  We chose the former option
for most of the analyses in this section.

\item Several {\it merging procedures} can be used.  One might begin by
merging the clusters with the nearest neighbors.  This is called Single Linkage 
clustering and is most familiar in astronomy where it is frequently called the
friends-of-friends algorithm. It tends to produce long stringy clusters, and is
equivalent to a  well-known divisive clustering procedure known as pruning the
minimal spanning tree.  Complete Linkage proceeds by maximizing the distance
between clusters, and leads to evenly bifurcating dendrograms.  For most of our
analysis, we choose Average  Linkage where the distance between two clusters
is the average of the distances between pairs of observations, where each
member of the pair comes from a different cluster.  This is a compromise
between Single and Complete Linkage and tends to give compact clusters.
Specifically, the distance between clusters K and L is given by (e.g. SAS
Institute Inc. 1989, pp. 529ff; Johnson \& Wichern 1992, pp. 584ff.)
\begin{equation}  
  D_{KL} = |{\overline{\bf x}_K-\overline{\bf x}_L}|^2 +                     
\frac{W_K}{n_K} + \frac{W_L}{n_L},
\end{equation}
where the bar indicates an unweighted mean, $W_K = \sum_{i=1}^{n_k} 
|{\bf x_i - \overline{x}}|^2$, and $n_k$ is the number of members of the
$k$-th cluster.  Another popular choice is Ward's minimum variance criterion
where the distance between the two clusters is the ANOVA (analysis of
variance) sum of squares between two clusters added up over all variables
(Ward 1963), \begin{equation}
    D_{KL} = |{\overline{\bf x}_K-\overline{\bf x}_L}|^2 /                       
(\frac{1}{n_K}+\frac{1}{n_L}).
\end{equation}
If the sample is generated by a mixture of multinormal (i.e. multidimensional
Gaussian) distributions where each distribution has covariance 
matrix of the form $\Sigma^2 I$, this method joins  clusters to maximize 
the likelihood at each level of the hierarchy,  and so is a special case of 
the model-based clustering methodology to be discussed in \S 5.

\item As the procedure gives a hierarchy from $N$ clusters with 1 object down
to one cluster with $N$ objects, the user must choose {\it how many clusters to
report} as scientifically important clusters.  This choice can be assisted by
examination of two statistics.  The  squared correlation coefficient, $R^2$,
states the fraction of the total variance  accounted for by a partition into $g$
clusters,
\begin{equation}
    R^2 = 1 - \frac{\Sigma_{j=1}^g W_j}{\Sigma^N_{i=1} 
     |{\bf x_i - \overline x}|^2}.
\end{equation}
The squared semi-partial correlation coefficient, $R^2_{sp}$ measures the
difference in the variance between the resulting cluster and the immediate
parent clusters normalized by the total sample variance,  \begin{equation}
      R^2_{sp} =  \frac{W_M - W_K -W_L} {\Sigma^N_{i=1} 
      |{\bf x_i  - \overline x}|^2}. 
\end{equation}
$R^2$ thus tells how much of the scatter is explained by a given level of
clustering,  and $R^2_{sp}$ tells how much improvement is achieved between
levels.   
\end{enumerate}

We emphasize again that there is no mathematically `best' choice, although
extensive experience with problems in many fields has led to a preference for
certain combinations (e.g. standardized variables and Ward's minimum variance
criterion).  We conducted extensive experiments with different choices.   

\subsection{Results}

The last several levels of the clustering tree for the 797 GRBs using the six
unit-free variables shown in Table 1, Average Linkage and a Euclidean metric
are shown in Figure 2 (left panel) with details in Table 3a.  The action taken at
each level is indicated in column 2 of Table 3, which may refer to a level higher
in the tree which (for brevity) is not shown here.  Two types of mergers are
seen: the incorporation of `twigs' of one or a few GRBs into a large preexisting
`trunk' (levels 1, 3, 4, 5 and 7); and the union of two substantial branches into a
single larger trunk (levels 2, 6 and 8). The first type has little effect on the
variance of the sample with $R^2_{sp} \leq 1$\%.  The single GRB brought
into the main trunk at level 1 is the distant outlier seen in several panels of
Figure 1.  The level 2 merger of clusters with 190 and 606 members is clearly
the most important structure, accounting for roughly 53\% of the variance of
the entire sample.  This is the bifurcation of the sample into two classes easily
seen in Figure 1 and noted by K93 and others. The principal finding that is not
immediately obvious from Figure 1 is the structure indicated at level 6.  The
main trunk of 599 bursts (plus a few twigs to be merged later) is divided into
groups of 93 and 506 bursts.  This division accounts for 10\% of the total
variance of the sample, indicated in both the $R^2$ and $R^2_{sp}$ values.  

\begin{figure}
\plotfiddle{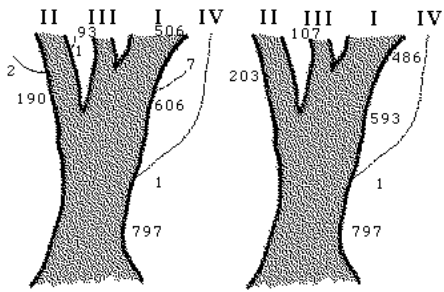}{1.5in}{0.0}{50}{50}{-100.}{000.}
\caption{Diagram of the base of the dendrogram of Average Linkage
hierarchical clustering procedures in 6 dimensions (left panel)  and 5
dimensions (right panel). The number of members in each branch is indicated
(see Table 3). Class IV is the spurious outlier.}
\end{figure}

We found that the twigs in the tree structure disappear if the peak flux
$P_{256}$ variable is omitted and the analysis is made in 5-dimensional space
(Figure 2, right panel, and Table 3b).  Here the largest cluster of 593 members
is formed by the union of clusters with 107 and 486 bursts, again accounting
for 10\% of the sample variance.  It is possible that $P_{256}$ is a nuisance
variable irrelevant to the basic astrophysics of GRBs, producing noisy `twigs'
seen in Table 3a and Figure 2 (left panel). 

We tested many variants of hierarchical clustering.  We replaced Average
Linkage hypothesis with Complete Linkage,  Single Linkage and Ward's
minimum variance criterion.  The Ward's criterion computation, for example,
gave three clusters  with 468, 145 and 184 bursts.  We clustered using
nonparametric density estimation based on the  100 nearest neighbors, and
clustered using the principal components rather than the  observed variables.
Various  methods were tried with both the observed $T_{90}$ values and
debiased $T_{90}^d$ values, with little effect on the results.  All methods
showed two strong clusters and the outlier but,  in some cases, the third cluster
appeared only weakly.   

\begin{figure}
\plotfiddle{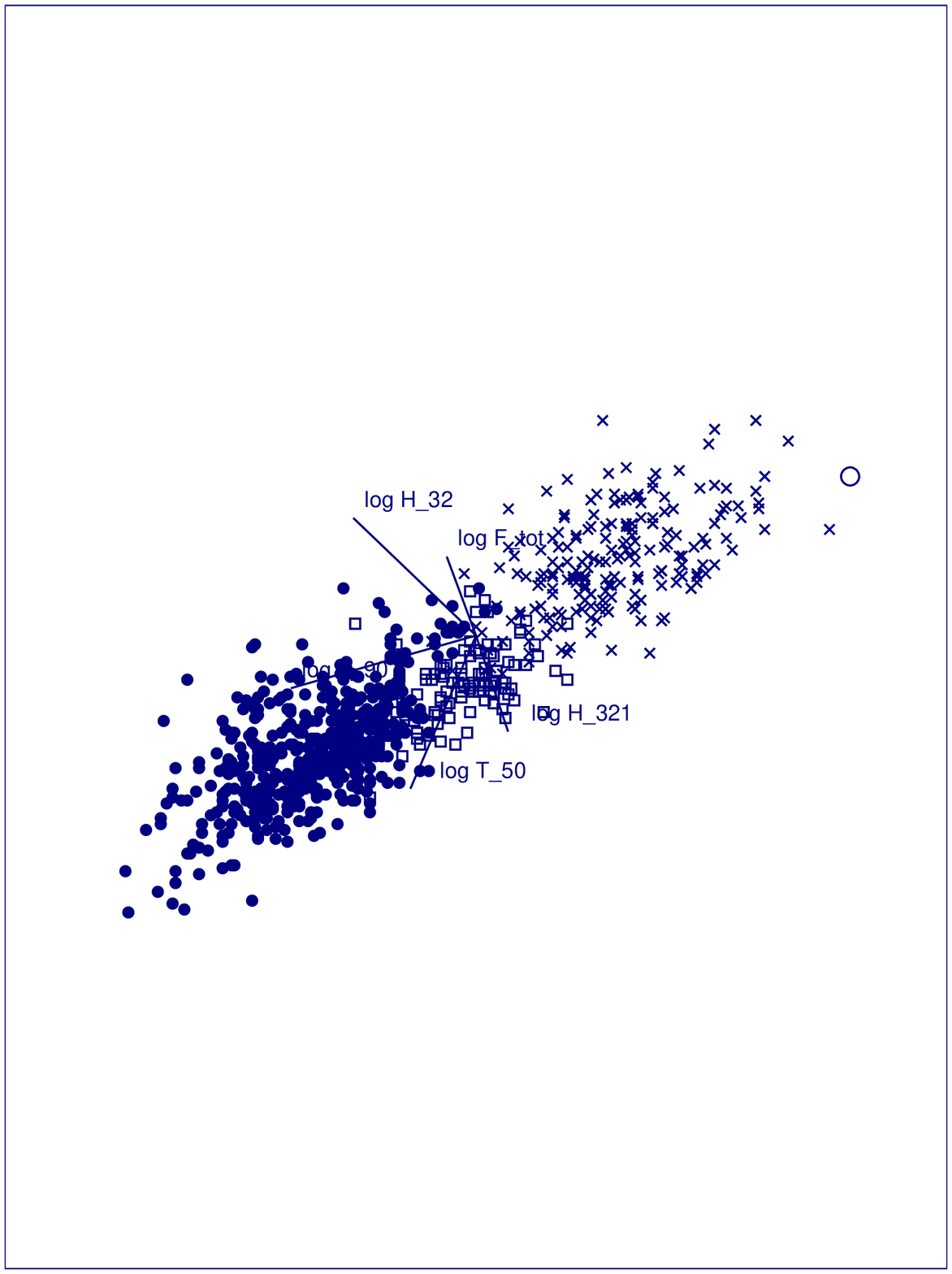}{3.5in}{0.0}{45}{45}{-270.}{-055.}
\plotfiddle{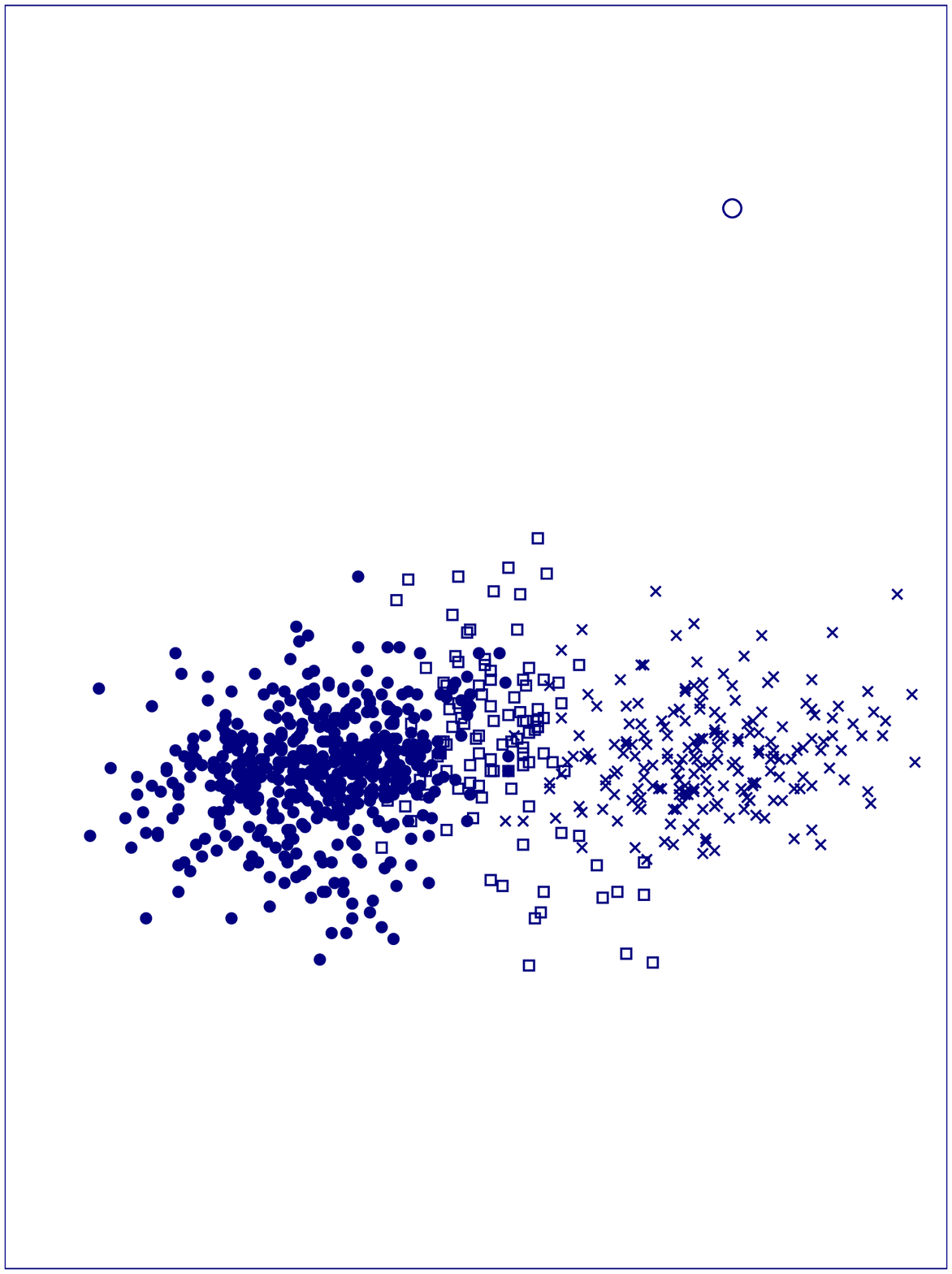}{3.5in}{0.0}{45}{45}{-020.}{210.}
\plotfiddle{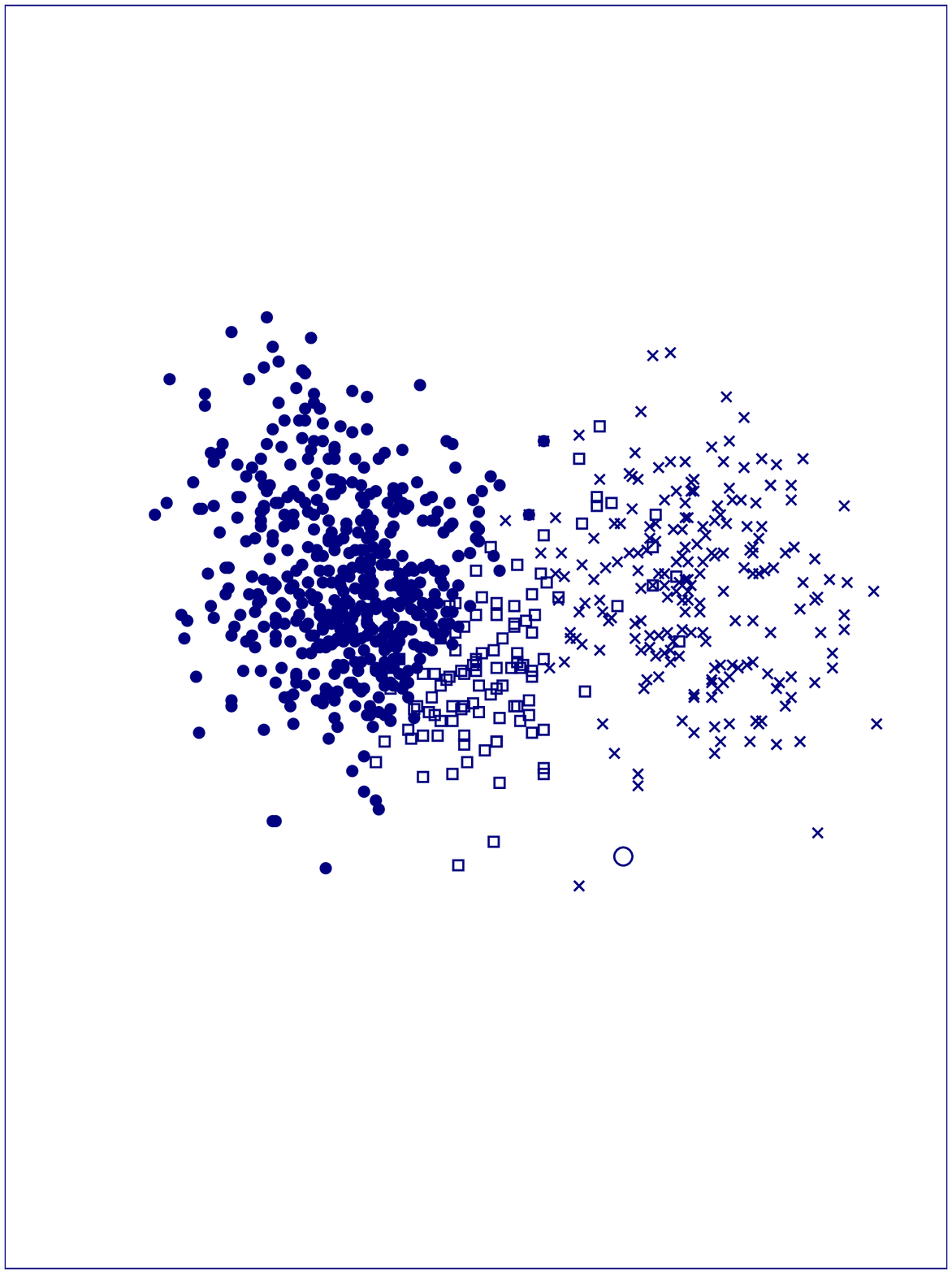}{3.5in}{0.0}{45}{45}{-270.}{185.}
\plotfiddle{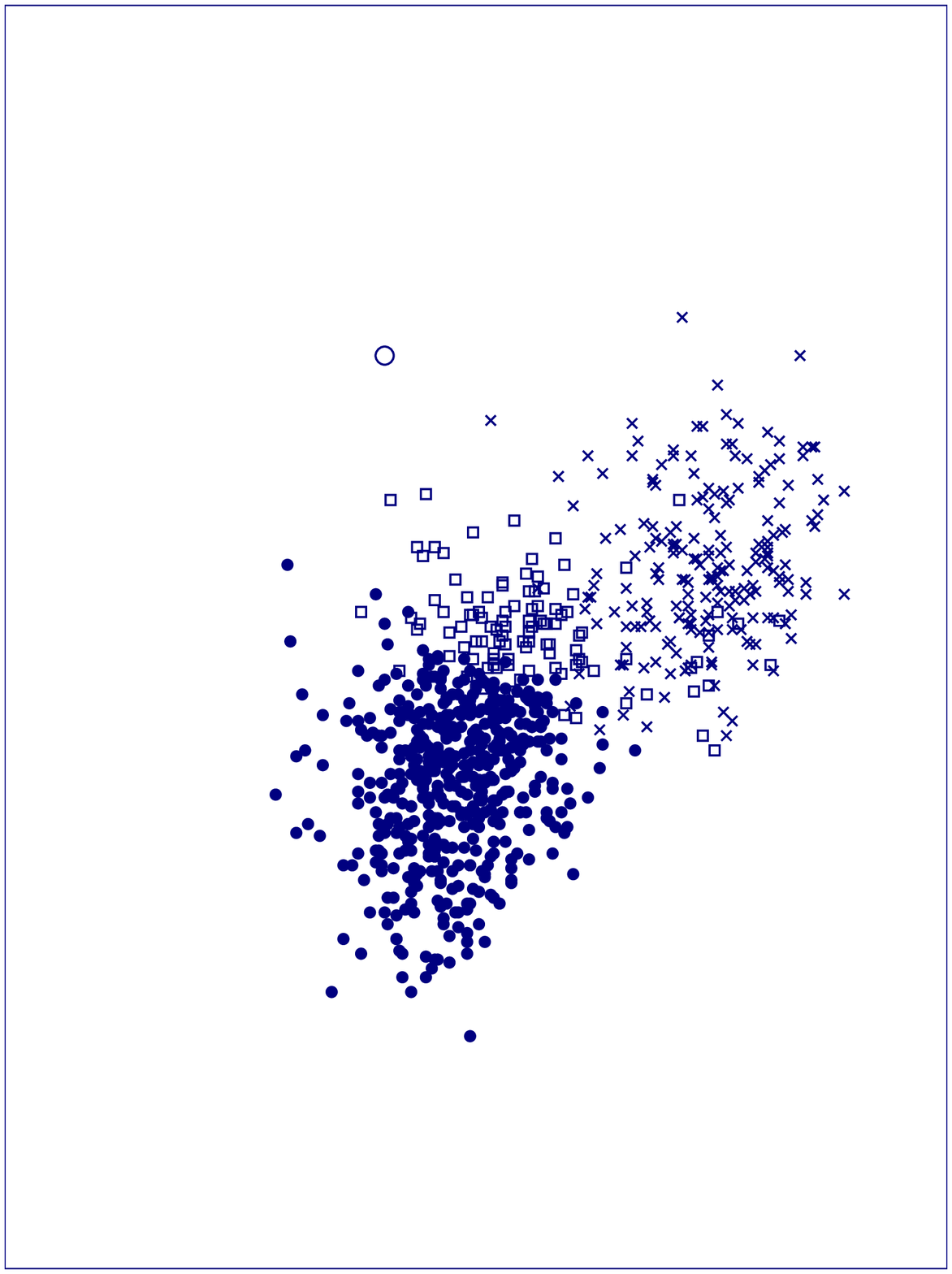}{3.5in}{0.0}{45}{45}{-020.}{450.}
\plotfiddle{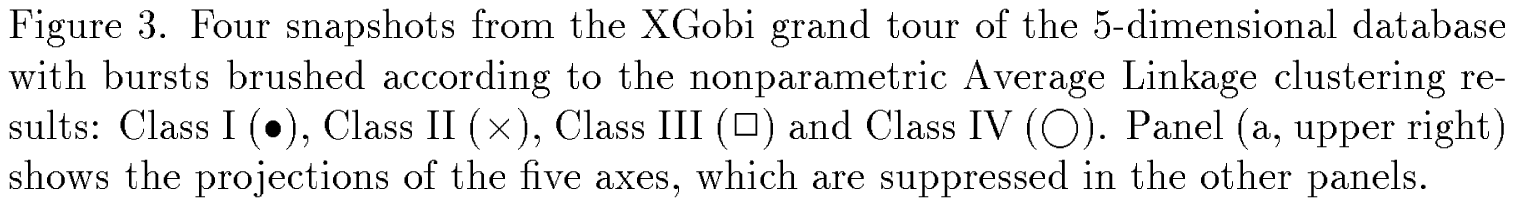}{0.5in}{0.0}{100}{100}{-300.}{-130.}
\caption{Four snapshots from the XGobi
grand tour of the 5-dimensional database with bursts brushed according to the
nonparametric Average Linkage clustering  results:  Class I ($\bullet$),  Class II
($\times$), Class III ($\Box$) and Class IV ($\bigcirc$). Panel (a) shows the
projections of the five axes, which are suppressed in the other panels.} 
\end{figure}

To proceed further, we choose a single clustering structure for detailed study:
the 5-dimensional Average Linkage analysis (Table 3b) with three clusters ---
Class I with 486 bursts, Class II with 203 bursts, Class III with 107 bursts. 
Class IV consisting of the single outlier is ignored because of independent
evidence that its properties are due to data of poor quality (\S 7). The
membership of these clusters is given in Table 4, and four projections of the
clusters onto two-dimensional scatter plots are shown in Figure 3.   These are
frames from the `grand tour' movie of the 5-dimensional dataset provided by
the XGobi software where each cluster is `brushed' with a different symbol. 
Note that, in general, there is no reason why classification structure should be
most evident in projections parallel to the variable axes shown in Figure 1.  It is
more important that the clusters show cohesion in many projections of the
dataset.  The grand tour of the 797 GRBs shows that Classes I, II and the
outlier are very distinct in most projections.  Class III often lies between Classes
I and II (e.g. top panels of Figure 3), but in other projections is offset from the
line between Classes I and II (e.g. bottom panels of Figure 3).  It also appears
elongated along some projections, while the larger Classes I and II appear
roughly hyperspherical.  
  
This analyses described here provide considerable evidence for three major
clusters and an outlier. But as some nonparametric clustering procedures did
not find a strong third cluster, there is some some worry that Class III is simply a
group of bursts with properties intermediate between the Classes I and II. 
While nonparametric hierarchical clustering methods can not address this
question, it can be investigated with parametric methods. 

\subsection{Validation of the classification}

Mathematically well-founded methods for evaluating the statistical significance
of a proposed multivariate classification scheme are available under the
assumption that the population is a multinormal mixture; that is, the objects of
each class are drawn from multivariate Gaussians.  All relationships between
the variables must thus be linear (as in Table 2), although the relationships may
differ between clusters.  There is no requirement of sphericity, so that clusters
may have shapes akin to pancakes or cigars with arbitrary orientations in 
multidimensional space.    The separate existence of each of the
postulated subpopulations can be tested using multivariate
analysis of variance (MANOVA).

The model can be expressed as follows (e.g. Johnson \& Winchern 1992, pp.
246ff).  For a $p$-dimensional dataset of $g$ clusters each with $n_l$
members, the $i$-th GRB in the $j$-th cluster gives a $p$-dimensional vector  
\begin{equation}
    {\bf X}_{ij} = {\bf  \mu} + {\bf \tau}_j + {\bf \epsilon}_{ij} 
\end{equation}
where ${\bf \mu}$ is the overall population mean, ${\bf \tau}_j$ is the offset
of the $j$-th cluster mean from ${\bf \mu}$, and ${\bf \epsilon}_{ij}$ are
independent normal variables with zero mean representing the scatter of
individual points about the mean.  We test the null hypothesis  
\begin{equation}
   H_0: {\bf \tau}_1 = {\bf \tau}_2 = \ldots = {\bf \tau}_g = 0  
\end{equation}
that the cluster means are not offset from each other.  We construct two
matrices of sums of squares and cross-products as follows:  
\begin{eqnarray}
   {\bf B} &=& \sum_{l=1}^g n_l (\overline{\bf x}_l-\overline{\bf x})               
                                (\overline{\bf x}_l-\overline{\bf x})^T \nonumber \\   
{\bf W} &=& \sum_{l=1}^g \sum_{j=1}^{n_l} (\overline{\bf
x}_{lj}-\overline{\bf x_l}) (\overline{\bf x}_{lj}-\overline{\bf x_l})^T, 
\end{eqnarray}
where $^T$ is the vector transpose.  Three test statistics have been proposed to
test the null hypothesis (e.g. SAS Institute Inc. 1989, pp. 17ff)):   
\begin{eqnarray}
     {\rm Wilks' ~Lambda} ~~~ \Lambda^* &=& {\rm det}({\bf W})/                
                                   {\rm det}({\bf B + W}), \nonumber \\
     {\rm Pillai's ~trace} ~~~  {\bf V}  &=& {\rm trace}[{\bf B(B+W)}^{-1}],    
                                   {\rm and} \\      
     {\rm Hotelling-Lawley's ~trace} ~~~ {\bf U} &=& {\rm trace} 
                                 ({\bf W}^{-1} {\bf B}).    \nonumber     
\end{eqnarray}
The distributions of these statistics have been determined mathematically. For
example, for large $N = \sum_{l=1}^g n_l$, the quantity $-(N-1-(p+g)/2) {\rm ln} 
\Lambda^*$ has approximately a chi-squared distribution with $p(g-1)$
degrees of  freedom (Wilks 1932; Bartlett 1938).  More generally, the
distributions are related  to the non-central $F$ distribution.  For the 2-sample
case, Hotelling-Lawley's trace is  commonly known as the Mahalanobis $D^2$
statistic.  One can thus accept or reject the null hypothesis that the clusters have
the same mean location at a chosen level of statistical significance.

The results of our MANOVA calculations are summarized in Table 5.  The
columns give the values of the three MANOVA statistics, followed by details
for the Wilks' $\Lambda^*$: the corresponding value of the $F$ statistic, the
numerator and denominator degrees of freedom for that $F$ value; and the
resulting $P$ value.  Details for Pillai's
and Hotelling-Lawley's traces are omitted, but give similar results in all cases. 
The first row tests the null  hypothesis that the Classes I, II and III have the
same mean, the second row tests the equality of Classes I and II, and so forth. 
The $F$ values are very high  in all cases,  indicating that the clusters are
different with extremely high statistical significance ($P << 10^{-4}$).\footnote{Note
that it is not meaningful to quote probabilities like $P = 10^{-8}$ as the tails of
the distribution are poorly determined unless the sample size is extremely large.}
This is a clear {\it quantitative} demonstration that at least two clusters exist
among GRBs (for a univariate test, see Ashman, Bird \& Zepf 1994), which was {\it qualitatively}
reported by Delazay et al. (1992) and K93. The other rows in Table 5 test the
hypotheses that each proposed class has the same mean as each other class.  

One problem with these MANOVA tests is that they are conditional on the
classification that has been found using the clustering algorithm.  Because the
clustering algorithm is constructed to find groups that are different from one
another, tests such as these tend to be biased towards finding structure, perhaps
where none exists.  Although the MANOVA results seem to indicate very
strong evidence of structure, they can not be taken as definitive for this reason. 
Tests arising from model-based clustering can overcome this problem, as
discussed in the following section.  

\section{Model-based maximum likelihood clustering analysis} 

\subsection{Methodological background}

In the previous section, we conducted a hierarchical clustering analysis without
making assumptions regarding the shapes of the clusters, but needed the
parametric assumption of normality to estimate the statistical significance of the
resulting classification.  It is reasonable to conduct the entire analysis, both
clustering and validation, within a model-based framework.  We report here an
analysis of this type again assuming that the GRB population consists of a
mixture of multivariate Gaussian classes.  Early development of this model for
clustering is discussed in McLachlan \& Basford (1988); we use more recent
developments here.  First, an initial classification for each possible number of
clusters is found via agglomerative hierarchical clustering (Murtagh \& Raftery
1984, Banfield  \& Raftery 1993,  Fraley 1998). Next, the EM
(Expectation-Maximization) algorithm is used to refine partitions obtained from
hierarchical clustering (Celeux \& Govaert 1995, Dasgupta \& Raftery 1998). 
Finally, the Bayesian Information Criterion is used to select the `best' partition
among those associated with different numbers of clusters (Dasgupta \&
Raftery 1998).

In the model considered here, the $p$-dimensional observations ${\bf x}_i$ are
drawn from $g$ multinormal groups, each of which is  characterized by a vector of parameters
${\bf \theta}_k$ for $k = 1, \ldots, g$.  Our goals are: to determine the number
of GRB types, $g$; to determine the cluster assignment of each burst; and
to estimate the mean ${\bf \mu}_k$ and covariance matrix ${\bf \Sigma}_k$ for
each cluster.  Following Fraley (1998), the density of an observation ${\bf
x}_i$ from the $k$th subpopulation is expressed as follows:
\begin{equation}
  f_k({\bf x}_i | \theta_k) \sim MVN({\bf \mu}_k, {\bf \Sigma}_k) ~~~~ 
                                             k = 1, \ldots, g 
  \end{equation}
where $MVN$ means multivariate normal.  
We estimate the parameters using the principle of maximum likelihood. In the
hierarchical clustering phase, we use the classification likelihood
\begin{equation}
L_{\scriptscriptstyle C}({\bf \theta, \gamma} \mid {\bf x}) =
\prod_{i=1}^N f_{\gamma_i}({\bf x}_i| \theta_{\gamma_i})
\end{equation}
where ${\bf x} = {\bf x}_1, {\bf x}_2, \ldots, {\bf x}_N$ represents the
observations and ${\bf \gamma} = \{\gamma_1, \gamma_2, \ldots, \gamma_N
\}$  is the cluster assignment: $\gamma_i = k$ when $x_i$ comes from the
$k$-th group.   Equivalently,
\begin{eqnarray}
\lefteqn{ L_{\scriptscriptstyle C}({\bf \mu}_1, \ldots, {\bf \mu}_k; 
       {\bf \Sigma}_1,\ldots,{\bf \Sigma}_k \mid {\bf x}) =} \nonumber \\
& &  \prod_{k=1}^g \prod_{i \in I_k} (2\pi)^{p/2} |{\bf \Sigma}_k|^{-1/2}
\exp \{-\frac{1}{2} ({\bf x}_i- {\bf \mu}_k)^T  {\bf \Sigma}_k^{-1} 
         ({\bf x}_i- {\bf \mu}_k) \},
\end{eqnarray}
where $I_k = \{i:\gamma_i = k\}$ is the set of indices corresponding to
observations belonging to the $k$-th group. 

The method used here for maximizing the likelihood (Fraley 1998) and
implemented in the {\it MCLUST} code involves parameterization of the ${\bf
\Sigma}_k$ matrices in terms of their eigenvectors and eigenvalues (analogous
to a principal components analysis), and iterative 
relocation of the clusters using the EM
Algorithm.  The EM (Estimation-Maximization) Algorithm (Dempster, Laird
\& Rubin 1977), one of the most successful methods in modern statistics, is a
procedure for iteratively maximizing likelihoods in a wide variety of
circumstances.   For example, the Lucy-Richardson algorithm in astronomical
image restoration is the EM Algorithm.  In the present application, we apply
EM to the mixture likelihood
\begin{equation}
L_{\scriptscriptstyle M}({\bf \theta, \gamma} \mid {\bf x}) =
\prod_{i=1}^N \sum_{k=1}^g \tau_k f_k({\bf x}_i| \theta_k); \ \ \ \
\sum_{k=1}^g \tau_k = 1,
\end{equation}
where $\tau_k$ are mixing probabilities associated with each group. For a given
number $g$ of components in the mixture, we use  EM to estimate the
conditional probability that observation ${\bf x}_i$ belongs to the $k$-th group
for each $i$ and selected $k$ via maximum likelihood.  Although the
computational procedure has some limitations (e.g. convergence of the EM
iterations is not guaranteed; clusters cannot be extremely small), it is generally
efficient and effective for Gaussian clustering problems when started from
reasonable partitions such as those produced by hierarchical agglomeration. 

We use the Bayes factor to assess the evidence for a given number of clusters
against a different number of clusters.  The Bayes factor, defined in the context
of Bayesian statistics, is the posterior odds for one model against the other
when the prior odds are equal to one (i.e., when one does not favor one model
over the other {\it a priori}).  Kass \& Raftery (1995) review the use of Bayes
factors in adjudicating between competing scientific hypotheses on the basis of
data.  The Bayes factor for a model $M_2$ against a competing model $M_1$
(say, for three $vs.$ two classes of GRBs) is defined as
\begin{equation}
    {\rm Bayes~ factor} = \frac{p({\bf x}|M_2)}{p({\bf x}|M_1)},
\end{equation}
where $p({\bf x}|M_j$) for $j = 1, 2$ is obtained by integrating the likelihood
times the prior density over the parameters of the model.  It can be viewed as a
likelihood ratio, but it differs from the usual frequentist ratio that underlies the
likelihood ratio test in that the latter is obtained by maximizing (rather than
integrating) the likelihood over the model parameters.  

Twice the logarithm of the Bayes factor can be approximated by the Bayesian
Information Criterion or BIC (Schwarz 1978),
\begin{equation}
    BIC = 2 (l_1 - l_2) - (m_1 - m_2) log N,
\end{equation}
where $l_1$ is the likelihood and $m_1$ is the number of parameters for one
mixture model, and similarly for $l_2$ and $m_2$. The BIC measures the
balance between the improvement in the likelihood and the number of model
parameters needed to achieve that likelihood.  While the absolute value of the
BIC is not informative, differences between the BIC values for two competing
models provide estimates of the evidence in the data for one model against
another.  Conventionally, BIC differences $< 2$ represent weak evidence,
differences between 2 and 6 represent positive evidence,  $6-10$ strong
evidence, and $> 10$ very strong evidence (Jeffreys 1961, Appendix B; Kass
\& Raftery 1995).   The use of the BIC in choosing clusters in a mixture or
clustering model is discussed by Roeder \& Wasserman (1997) and Dasgupta
\& Raftery (1998).  

Bayes factors and BIC have the advantage that they can be used to assess the
evidence {\it for} a null hypothesis, unlike standard significance tests which can
only reject a null hypothesis.  They can also easily be used to compare 
non-nested models, again unlike standard significance tests which require
competing models to be nested. 

\subsection{Results and validation}

To reduce the dimensionality of the problem and the complexity of the
calculation,  we eliminated the highly redundant $T_{50}$ and $H_{32}$
variables (see Figure 1)  and considered only the 3 variables $T_{90}$,
$F_{tot}$ and $H_{321}$ for the  sample of 797 BATSE GRBs.  The {\it
MCLUST} model-based clustering procedure described  above was run for
trials of $g = 1, 2,  \ldots, 24$ groups.  The resulting values of 
BIC($g$) are plotted in Figure 4.  The
maximum BIC is achieved for three classes.  Most importantly, the BIC value
for $g=3$ is $\simeq 68$ units above that for $g=2$.  This corresponds to 
strong evidence indeed for the presence of three groups rather than two.  This
result strongly confirms the analysis in \S 4 indicating the existence of three
clusters, and this time is free of the problem that the MANOVA tests are
conditional on the estimated partition.  The result here takes account of 
the fact that the partition is not known in advance. 

We have also calculated the BIC for $g=1, \ldots, 9$ with various constraints
on the covariance matrix {\bf $\Sigma$} such as hypersphericity and uniformly
shaped ellipsoids.  Spherical clusters give poor fits.  Uniform ellipsoids  give
good fits with 4 and 8 clusters.  But in all cases, the maximum likelihood 
assuming 2 clusters is much lower than the likelihood of $\geq 3$ clusters. 

The cluster assignment vector ${\bf \gamma}$ for the $g=3$ model with
unconstrained {\bf $\Sigma$} is given in Table 4. Over 85\% of the
assignments are the same as those obtained from the nonparametric hierarchical
clustering procedure in \S 4, so that we note only  differences between the two
clustering results with $^*$ and $^{\dag}$ markings.   All but one of the 96
assignment differences move bursts from Classes I and II into Class III.  The
close agreement between the cluster assignments in the two methods 
reinforces confidences in the conclusions from both of them. 

\begin{figure}
\plotfiddle{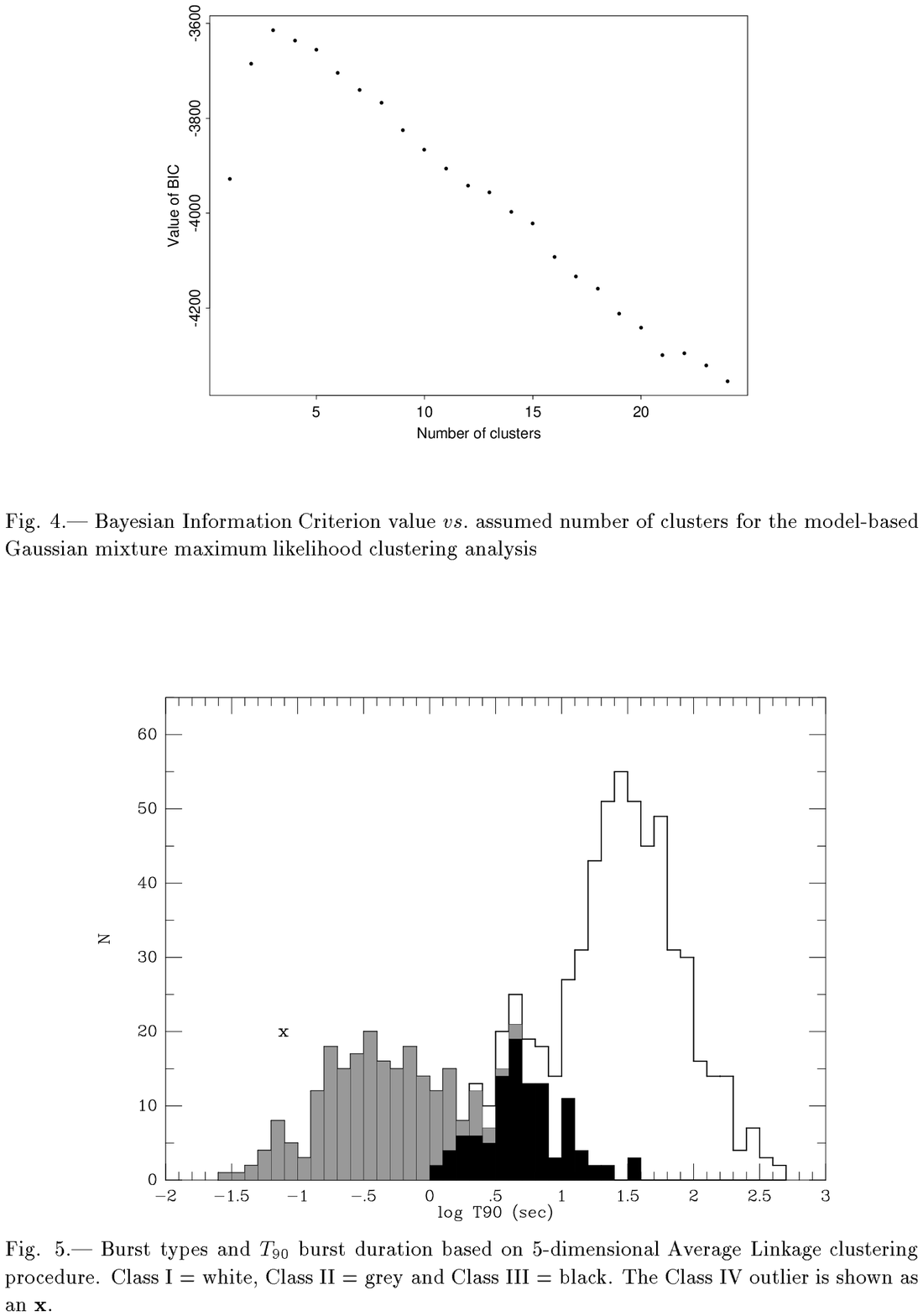}{10.0in}{0.0}{100.}{100.}{-300.}{050.}
\end{figure}
\newpage

\section{Cluster properties}

We can now examine the properties of GRBs within each cluster with
reasonable confidence that the populations are distinct from each other but
internally homogeneous.  These properties become inputs to astrophysical
theories seeking to explain GRB bulk properties.  Table 6a lists the means and
standard deviations of the principal variables for each cluster based on both the 
nonparametric and model-based clustering procedures.  The two methods give
very similar results.  The three types are well-separated in the burst duration
variables: Cluster I bursts have the longest durations around $10$-$20$
seconds, Cluster II bursts have the shortest durations below 1 second, and
Cluster III bursts have intermediate durations around 2 seconds. This is shown
clearly in Figure 5, which projects each class onto the univariate $T_{90}$
axis.  Cluster III bursts are also intermediate in their fluences, although their
fluence distribution overlaps that of the fainter Class II bursts.  The hardness
ratios of all three clusters overlap considerably, but Class III bursts have the softest
spectra and Class I  bursts have intermediate spectra.   We can thus classify the
types in the three principal dimensions Duration/Fluence/Spectrum (Table 6c):
Class I is long/bright/intermediate, Class II is short/faint/hard, and Class III is
intermediate/intermediate/soft.

A major constraint for the astrophysical interpretation of GRBs has been the
remarkable isotropy of their spatial distribution in the celestial sphere.  It is
possible that, while the bulk of GRBs are isotropic and have an inferred
extragalactic origin, some class of GRBs have significant anisotropy which
would reflect a Galactic origin (see Lamb 1995).  We apply four statistical tests
for isotropy  discussed by Briggs (1993) and applied by Briggs et al. (1996) to
various subsamples of the 3rd BATSE Catalog of GRBs.  The statistics are:
$<$cos$\theta>$, where $\theta$ is the angle between a burst and the Galactic
center; $<$sin$^2 b - \frac{1}{3}>$, where $b$ is the Galactic latitude;
Rayleigh-Watson $\cal W$; and Bingham $\cal B$.  $<$cos$\theta>$ tests the
dipole moment around the Galactic center, $<$sin$^2 b - \frac{1}{3}>$ tests
the quadrapole moment with respect to the Galactic plane, $\cal W$ tests the
dipole moment around any point in the celestial sphere, and $\cal B$ tests the
quadrapole moment around any plane or two poles. The expected values for
the four statistics assuming random isotropic distribution on the sphere are 0, 0,
3 and 5 respectively.  The asymptotic distributions of these statistics are known. 

Table 6b shows the results of this analysis for Clusters I-III, kindly calculated
for us by Michael Briggs.  No deviations from isotropy are found.  The
$<$cos$\theta>$ and $<$sin$^2-\frac{1}{3}>$ values lie within one standard
deviation of the expected value for a random distribution.  The $\cal W$ and
$\cal B$ values must be larger than the expected value to indicate anisotropy. 
The only such case, Class II with $\cal B =$7.32, has a deviation with very low
significance ($Prob < 0.2$).  We thus do not confirm Belli's (1997) report of
significant differences in spatial distributions of burst Classes I and III, although
we did not specifically test the Galactic latitude distribution.   

In principle, the relative populations of the three classes may be an important
constraint on astrophysical theory.  We find that Class I contains more than half
of the bursts with the remainder divided between Class II and Class III (Table
6c).  But we do not believe our analysis gives a precise census for two reasons. 
First, the exact assignments of individual bursts to clusters depend on the
detailed assumptions of the clustering algorithms.  For example, Class II is
larger than Class III in the 5-dimensional  nonparametric procedure but is
smaller in the 3-dimensional model-based procedure.  Second, the numbers of 
weaker bursts in Classes II and III are strongly dependent on the details of the
BATSE instrument's burst triggering process which produces a complicated
truncation bias for fainter bursts.

We look for structure within each of the clusters by computing correlation
coefficients similar to those in \S 3 for the entire sample.  Results are given in
Table 7.  Here we see a systematic difference between the two clustering
methodologies: nonparametric Average Linkage clustering tends to give
stronger correlations between the variables than the model-based clustering. 
For example, in the nonparametric analysis we find significant positive correlations
between total fluence and hardness in Classes I and II, and a correlation between
duration and fluence in Class II.  However, we
attach more credence to the model-based results for this purpose than the
Average Linkage results because the former method is specifically designed to
provide optimal estimates of the within-group covariances given the clustering
model.  The model-based results do not give strong evidence for any non-zero
correlations between variables, suggesting that the partition into three clusters
explains all of the correlation between variables in the full dataset.  

\section{Discussion}

We thus find, using clustering and validation methods with different
mathematical underpinnings, that three classes of GRBs are present in our large
subset of the Third BATSE Catalog.  Most of the structure can be found using
three fundamental burst properties, Duration/Fluence/Spectrum.  The class
properties and relation to previous research can be briefly summarized as
follows: 
\begin{description}

\item[Class I] These long/bright/intermediate bursts correspond to the 
well-known populous long-soft class of K93 and others.  Within this group, we
do not confirm a hardness-duration correlation reported by Dezalay et al.
(1996) and Horack \& Hakkila (1997).  

\item[Class II] This short/faint/hard group corresponds to the short-hard burst
type of K93 and others.  Fluence-duration and fluence-hardness correlations
may tentatively be present within the class.  Note that while the mean location
of this type is consistent in the two clustering schemes, its size and population
(e.g. 1/2 or 1/4 that of Class I) differs between clustering algorithms.  

\item[Class III] The discovery of this group with intermediate/intermediate/soft
properties is the principal result of this study.  The group is easily distinguished
in the projections of Figure 3, but can also be discerned in some panels of
Figure 1.    For example, it lies between Classes I and II in the
$T_{50}-H_{32}$, $T_{90}-F_{tot}$ and $T_{90}-H_{321}$ scatter plots. In
the univariate $T_{90}$ distribution shown in Figure 5,  Class III accounts for
most, but not all, of the bursts in the small peak around $2 < T_{90}<5$ sec
between the major short and long duration peaks.  It is possible that our Class
III is related to the class of no-high-energy (NHE) burst and peaks discussed by
Pendleton et al. (1997).  These bursts have unusually weak $F_4$ emission,
soft $50-300$ keV spectra, and low $F_{tot}$.  However, the NHE class does
not appear to exhibit a clear duration segregation from other bursts as we find
for Class III.  Class III does not appear to be the third cluster found by
Baumgart (1994, see his Table 3), but the high dimensionality of his analysis
prevents a simple comparison with our low dimensionality study.  

\item[Outlier]  BATSE trigger event 2757, burst 3B 940114, is the outlier in
the nonparametric analysis of \S 4 and is clearly visible in many projections in
Figures 1 and 3.\footnote{The model-based analysis of \S 5 cannot locate clusters with
very few members and assigned this event to Class II.  An extension of
model-based clustering that models outliers as Poisson noise can do this
(see Banfield and Raftery 1993, Dasgupta and Raftery 1998), but it does not seem 
necessary in this application.} It has an exceedingly
soft hardness ratio and short burst duration.  But examination of the original
BATSE database shows that the $F_1-F_4$ fluxes are very weak with large
measurement uncertainties.  The published 3B catalog gives only an upper limit
to its total fluence and no estimate of its hardness ratio (Meegan et al. 1996).  
The unusual properties of this burst are thus illusory and are due to its very
weak fluence. 
\end{description}

The multivariate analysis described here is not comprehensive and may  not
have uncovered all of the structure in the Third BATSE Catalog of bulk GRB
properties.  Our reduction of dimensionality may have been too severe,  omitting,
for example, the potentially important $F_4$ as a  distinct variable (Pendleton
et al. 1997; Bagoly et al. 1997).  Many  methodological options were not
exercised.  For example, it would be valuable to repeatedly apply the
$k$-means  partitioning algorithm to the database under the assumption that 
three clusters are present (see Murtagh 1992 for an astronomical application of
this method), check for skewness or kurtosis in the clusters, and undertake an
oblique decision tree analysis to give analytical formulation to hyperplanes
separating the clusters (see White 1997).  Codes for these and many other
multivariate techniques are publicly available through the Web metasite {\it
StatCodes} at www.astro.psu.edu/statcodes.  

However, the efforts described here are far more capable of finding and
quantifying clustering in the database than most previous analyses (\S 1). 
Previous studies have been based on qualitative rather than quantitative
procedures for identifying structures, and provide no statistical validation of
their claims.  It is thus not surprising that we uncovered structure  missed by
previous researchers. In particular, our confidence in the presence of a third
cluster, Class III,  is strong.  Two completely independent mathematical 
procedures (\S 4 and \S 5) found very similar structure, each validated with
high statistical confidence.  

It is possible that the clustering reported here is indeed present in the database,
but does not have an astrophysical origin.  The complex triggering mechanism
of the BATSE instrument mechanism and biases in bulk property values at low
signal-to-noise ratio are two problems that probably affect the multivariate
structure.  We have investigated one manifestation of the latter effect using
$T_{90}^d$ and found no effect on our results.   Instrumental biases generally
affect the number of bursts found in some regions of the multivariate
hyperspace (thereby biasing log $N$ - log $S$ distributions)  and may alter the
location of clusters, but are unlikely to cause the appearance of clustering that is
not present in the underlying population.  Nonetheless, since the BATSE
instrument identifies bursts on three separate timescales, it is possible that the
third cluster here is related to a selection effect associated with the BATSE
triggering mechanisms.  

We conclude that the Third BATSE Catalog shows three statistically significant
types of bursts (Duration/Fluence/Spectrum):  Class I GRBs are
long/bright/intermediate, Class II GRBs are short/faint/hard, and Class III GRBs
are intermediate/intermediate/soft.  Unless the separation of Class III from the
other types is due to some subtle BATSE instrumental effect, these types are
likely to be real and their existence should be considered an important input into
astrophysical theories for GRBs.  For example, the three types may reflect different
types of external environments and internal shocks in relativistic fireball models
(M\'esz\'aros \& Rees 1993; Panaitescu \& M\'esz\'aros 1998).  Note that
statistical anlaysis is unable to determine whether burst types represent fundamentally 
different astrophysical processes or distinct conditions within a single astrophysical model.  

Our results can be confirmed and extended in two fashions.  First, the analysis
described here can be validated with several hundred more bursts collected by
BATSE since the September 1994 cutoff in the database used here.  Second,
following Baumgart (1994), the dimensionality of the problem can be enlarged
to include detailed characteristics of the burst temporal behaviors.  Burst
smoothness $vs.$ peakiness, characteristic wavelet scales, spectral evolution,
and other parameters can be included.  With this enlarged database, one can
perform both an unsupervised exploratory cluster analysis similar to that
described here, and MANOVA-type analyses that assume the existence of the
three groups to determine whether the clusters have distinctive temporal
properties.  

\acknowledgements

This work was supported by NASA grant NAGW-2120 and NSF grant 
DMS-9626189 funding astrostatistical research at Penn State, and ONR grants
N-00014-96-1-0192 and N-00014-96-1-0330 funding statistical research at
Washington.  SM would like to thank the Departments of Astronomy \&
Astrophysics and Statistics at Penn State for hospitality while this work was
underway.  We are very grateful to Jay Norris (NASA-GSFC) for providing
debiased duration values and Michael Briggs (NASA-MSFC) for  performing
isotropy calculations. We also thank Drs. Briggs, Norris and Peter M\'esz\'aros
(Penn State) for sharing many valuable insights into GRB astronomy.

\newpage



\newpage

\begin{table*}
\caption{Average GRB properties for the entire sample}
\begin{tabular}{lrc}
&&\\ \hline
\multicolumn{1}{c}{Variable}               &  Mean & S.D$^a$  \\ \hline  
log $T_{50}$ (sec)                         &  0.55 & 0.92 \\  
log $T_{90}$ (sec)                         &  0.96 & 0.92 \\  
log $F_{tot}$ (erg cm$^{-2}$)              & -5.61 & 0.76 \\  
log $P_{256}$ (photons s$^{-1}$ cm$^{-2})$ &  0.16 & 0.45 \\  
log $H_{321}$                              &  0.25 & 0.33 \\  
log $H_{32}$                               &  0.48 & 0.30 \\  \hline 
\multicolumn{3}{l}{$^a$  Sample standard deviation}
\end{tabular}
\end{table*}

\begin{table*}
\caption{Correlation coefficients for the entire sample}
\begin{tabular} {lrrrrrr}
&&&&&& \\ \hline
               & log $T_{50}$ & log $T_{90}$ & log $F_{tot}$ &
log $P_{256}$ & log $H_{321}$ & log $H_{32}$   \\ \hline  
log $T_{50}$   &  1.00 &       &       &       &       &       \\  
log $T_{90}$   &  0.97 &  1.00 &       &       &       &       \\  
log $F_{tot}$  &  0.63 &  0.66 &  1.00 &       &       &       \\  
log $P_{256}$  & -0.01 &  0.04 &  0.59 &  1.00 &       &       \\  
log $H_{321}$  & -0.36 & -0.36 &  0.02 &  0.24 &  1.00 &       \\  
log $H_{32}$   & -0.35 & -0.35 & -0.00 &  0.19 &  0.96 &  1.00 \\  \hline 
\end{tabular}
\end{table*}

\begin{table*}
\caption{Average linkage hierarchical cluster analysis}
\begin{tabular} {rcrrr}
&&&& \\ \hline
Level & Merger & Members & $R^2_{sp}$ & $R^2$ \\ \hline
\multicolumn{5}{c}{(a) Six-dimensional analysis} \\   
8   & 10 + 15     & 506  & 0.08 & 0.65  \\ 
7   & 14 + 137    &  93  & 0.00 & 0.65  \\ 
6   &  8 + 7      & 599  & 0.10 & 0.55  \\ 
5   &  9 + 266    & 188  & 0.00 & 0.55  \\ 
4   &  5 + 26     & 190  & 0.00 & 0.55  \\ 
3   &  6 + 12     & 606  & 0.01 & 0.54  \\ 
2   &  3 + 4      & 796  & 0.53 & 0.01  \\ 
1   &  2 + 616    & 797  & 0.00 & 0.00  \\ 
&&&& \\
\multicolumn{5}{c}{(b) Five-dimensional analysis} \\   
6   & 15 + 21     & 107  & 0.01  & 0.70  \\ 
5   & 10 + 8      & 486  & 0.01  & 0.69  \\ 
4   &  7 + 20     & 203  & 0.01  & 0.68  \\ 
3   &  6 + 5      & 593  & 0.10  & 0.58  \\ 
2   &  3 + 4      & 796  & 0.58  & 0.00  \\  
1   &  2 + 616    & 797  & 0.01  & 0.00  \\ \hline
\end{tabular}
\end{table*}

\newpage

\begin{figure}
\plotfiddle{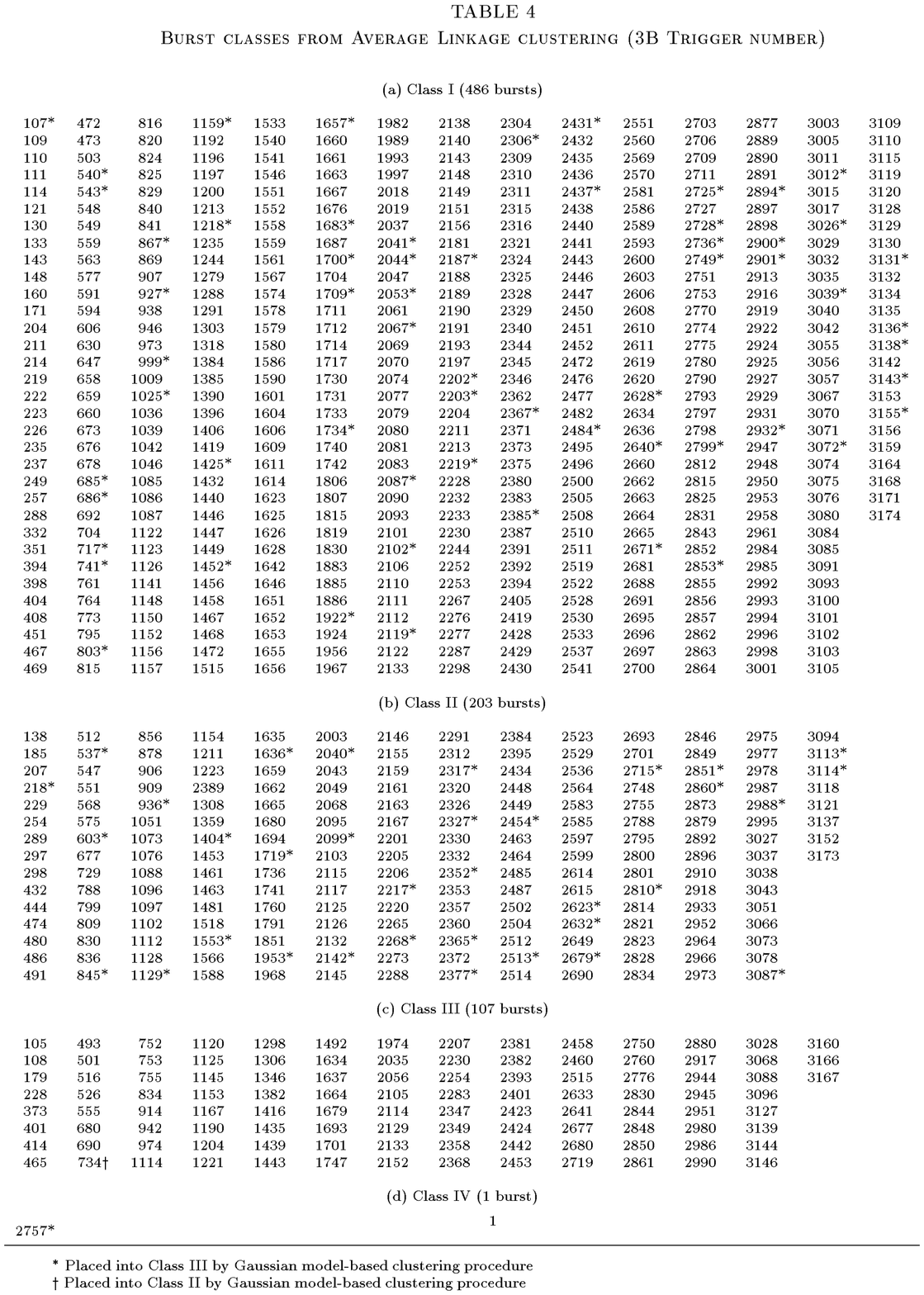}{10.0in}{0.0}{100.}{100.}{-300.}{050.}
\end{figure}

\begin{table*}
\tablenum{5}
\caption{Multivariate analysis of variance statistical tests} 
\begin{tabular}{lrrrrrrr}
&&&&&&& \\ \hline
Classes & Wilks' & Pillai's & Hotelling's & \multicolumn{4}{c}{Wilks'
$\Lambda^*$} \\ \cline{5-8}
  & $\Lambda^*$ & Trace & Trace & F & Num. dof & Den. dof & Prob. \\  \hline
I, II, III & 0.153 & 0.934 & 4.96 & 245. & 10 & 1578 & $<$0.0001 \\ 
I, II      & 0.159  & 0.840 & 5.27 & 722. &  5 & 683 & $<$0.0001 \\
I, III     & 0.515 & 0.485 & 0.94 & 111. &  5 & 587 & $<$0.0001 \\
II, III    & 0.301 & 0.699 & 2.32 & 141. & 5  & 304 & $<$0.0001 \\ \hline
\end{tabular}
\end{table*}

\newpage

\begin{table*}
\tablenum{6}
\caption{Class properties}  
\begin{tabular}{lcrrr}
&&&& \\ \hline
Variable & Method$^1$ & \multicolumn{3}{c}{Class} \\ \cline{3-5}
&& \multicolumn{1}{c}{I} & \multicolumn{1}{c}{II} &
\multicolumn{1}{c}{III} \\ \hline
\multicolumn{5}{c}{(a) Means and standard deviations} \\
log $T_{50}$  & NP &  1.13 $\pm$ 0.44 &  -0.80 $\pm$ 0.41 & 0.33 $\pm$ 0.27 \\
 log $T_{90}$  & NP     &  1.55 $\pm$ 0.40 &  -0.42 $\pm$ 0.44 & 0.71 $\pm$ 0.32 \\
            & MB &  1.22 $\pm$ 0.39 &  -0.91 $\pm$ 0.35 & 0.29 $\pm$ 0.41 \\
log $F_{tot}$ & NP     & -5.21 $\pm$ 0.59 & -6.37 $\pm$ 0.57 & -6.11 $\pm$ 0.37 \\
              & MB     & -5.13 $\pm$ 0.58 & -6.46 $\pm$ 0.54 & -5.93 $\pm$ 0.47 \\
log $H_{321}$ & NP     &  0.19 $\pm$ 0.27 &  0.51 $\pm$ 0.27 & 0.08 $\pm$ 0.40 \\ 
              & MB     &  0.21 $\pm$ 0.26 &  0.52 $\pm$ 0.28 & 0.16 $\pm$ 0.40 \\
log $H_{32}$  & NP     &  0.43 $\pm$ 0.23 &  0.70 $\pm$ 0.26 & 0.35 $\pm$ 0.39 \\ 
&&&& \\
\multicolumn{5}{c}{(b) Isotropy} \\
$<$cos $\theta$$>$  & NP &  0.015 & -0.041 & 0.010 \\
$<$sin$^2~b - 1/3>$ & NP & -0.012 & -0.025 & 0.028 \\
Rayleigh-Watson     & NP &  0.39  &  1.28  & 1.80  \\
Bingham             & NP &  2.02  &  7.32  & 1.95  \\
&&&& \\
\multicolumn{5}{c}{(c) Summary} \\ 
Number   & NP &       486    &       203    &       107    \\
         & MB &       426    &       170    &       201    \\
Duration &    &       long   & short  & intermediate \\
Fluence  &    &       bright & faint      & intermediate  \\
Spectrum &    & intermediate &       hard   &     soft     \\ \hline
\multicolumn{5}{l}{\hspace*{-2mm}$^1$ NP = nonparametric clustering
analysis in 5 dimensions (\S 4)} \\
\multicolumn{5}{l}{\hspace*{1mm}MB = model-based clustering analysis in 3
dimensions (\S 5)} 
\end{tabular}
\end{table*}

\begin{table*}
\tablenum{7}
\caption{Correlation coefficients within classes}
\begin{tabular}{lrrrrr}
&&&&& \\ 
& log $T_{50}$ & log $T_{90}$ & log $F_{tot}$ & log $H_{321}$ & log $H_{32}$   \\
\hline
\multicolumn{6}{c}{(a) Class I -- Nonparametric clustering (N=486)}\\
\multicolumn{6}{c}{$|r| > 0.15$ corresponds to $P < 0.001$ significance level}
\\
log $T_{50}$  &  1.00 &       &       &       &       \\
log $T_{90}$  &  0.88 &  1.00 &       &       &       \\
log $F_{tot}$ &  0.10 &  0.22 &  1.00 &       &       \\
log $H_{321}$ & -0.11 & -0.08 &  0.39 &  1.00 &       \\
log $H_{32}$  & -0.11 & -0.08 &  0.38 &  0.97 &  1.00 \\
&&&&& \\
\multicolumn{6}{c}{(b) Class I -- Model-based clustering (N=426)} \\
\multicolumn{6}{c}{$|r| > 0.16$ corresponds to $P < 0.001$ significance level}
\\
log $T_{50}$  &  1.00 &       &       &       &       \\
log $T_{90}$  &   N/A &   N/A &       &       &       \\
log $F_{tot}$ &   N/A &  0.01 &  1.00 &       &       \\
log $H_{321}$ &   N/A & -0.01 &  0.06 &  1.00 &       \\
log $H_{32}$  &   N/A &   N/A &   N/A &   N/A &   N/A \\
&&&&& \\
\multicolumn{6}{c}{(d) Class II -- Model-based clustering (N=170)} \\
\multicolumn{6}{c}{$|r| > 0.25$ corresponds to $P < 0.001$ significance level}
\\
log $T_{50}$  &  1.00 &       &       &       &       \\
log $T_{90}$  &   N/A &   N/A &       &       &       \\
log $F_{tot}$ &   N/A & -0.03 &  1.00 &       &       \\
log $H_{321}$ &   N/A & -0.05 &  0.02 &  1.00 &       \\
log $H_{32}$  &   N/A &   N/A &  N/A  &  N/A  &   N/A \\
&&&&& \\
\multicolumn{6}{c}{(e) Class III -- Nonparametric clustering (N=107)} \\
\multicolumn{6}{c}{$|r| > 0.32$ corresponds to $P < 0.001$ significance
 level}
\\
log $T_{50}$  &  1.00 &       &       &       &       \\
log $T_{90}$  &  0.86 &  1.00 &       &       &       \\
log $F_{tot}$ &  0.02 &  0.06 &  1.00 &       &       \\
log $H_{321}$ & -0.24 & -0.34 &  -0.16 &  1.00 &       \\
log $H_{32}$  & -0.22 & -0.32 &  -0.22 &  0.95 &  1.00 \\
&&&&& \\
\multicolumn{6}{c}{(f) Class III -- Model-based clustering (N=201)} \\
\multicolumn{6}{c}{$|r| > 0.23$ corresponds to $P < 0.001$ significance  level} \\
log $T_{50}$  &  1.00 &       &       &       &       \\
log $T_{90}$  &   N/A &   N/A &       &       &       \\
log $F_{tot}$ &   N/A &  0.03 &  1.00 &       &       \\
log $H_{321}$ &   N/A & -0.01 &  0.07 &  1.00 &       \\
log $H_{32}$  &   N/A &   N/A &   N/A &   N/A &   N/A \\ \hline
\end{tabular}
\end{table*}


\newpage

\begin{references}

Akaike, H. 1979, Biometrika, 66, 237

Ashman, K., Bird, C. M. \& Zepf, S. E. 1994, AJ, 108, 2348 

Babu, G. J. \& Feigelson, E. D. 1996, Astrostatistics (London:Chapman \&
Hall) 

Bagoly, Z., M\'esz\'aros, A., Horv\'ath, I., Bal\'azs, L. G. \& M\'esz\'aros, P.
1997, ApJ, submitted 

Banfield, J. D. \& Raftery, A. E. 1993, Biometrics, 49, 803

Bartlett, M. S. 1938, Proc. Camb. Phil. Soc., 34, 33

Baumgart, C. W. 1994, in Applications of Artificial Neural Networks V, edited
by S. K. Rogers \& D. W. Ruck, SPIE Proc. vol. 2243, (Bellingham
WA:SPIE), 552 

Belli, B. M. 1997, ApJ, 479, L31

Beyer, W. H. (editor) 1968, Handbook of Tables for Probability and Statistics, 
2nd ed. (Boca Raton FL:CRC Press)

Bhat, P. H., Fishman, G. J., Meegan, C. A., Wilson, R. B., Kouveliotou, C.,
Paciesas, W. S., Pendleton, G. N. \& Schaefer, B. E. 1994, ApJ, 426, 604   
Briggs, M. S. 1993, ApJ, 407, 126

Briggs, M. S., Paciesas, W. S., Pendleton, G. N., Meegan, C. A., Fishman, G.
J., Horack, J. M., Brock, M. N., Kouveliotou, C., Hartmann, D. H. \&
Hakkila, J. 1996, ApJ, 459, 40

Celeux, G. \& Govaert, G. 1995, Pattern Recognition, 28, 781  

Dasgupta, A. \& Raftery, A. E. 1998 , J. Amer. Stat. Assn. in press  

Dempster, A. P., Laird, N. M. \& Rubin, D. B. 1977, J. Royal Stat. Soc. B.,
31, 1

Dezalay, J.-P., Barat, C., Talon, R., Sunyaev, R., Terekhov, O. \& Kuznetsov,
A. 1992, in Gamma-Ray Bursts, edited by W. S. Paciesas \& G. J. Fishman,
AIP Conf. 265, (NY:AIP), 304

Dezalay, P.-P., Lestrade, J. P., Barat, C., Talon, R., Sunyaev, R., Terekhov, O.
\& Kuznetsov, A. 1996, ApJ, 471, L27

Feigelson, E. D. \& Babu, G. J. 1997, in New Horizons from Multiwavelength
Sky Surveys, edited by H. MacGillivray and B. Lasker, IAU 179,
Dordrecht:Kluwer, in press

Fenimore, E. E., in't Zand, J. J., Norris, J. P., Bonnell, J. T. \& Nemiroff, R. J.
1995, ApJ, 448, L101

Fishman, G. J. \& Meegan, C. A. 1995, ARAA, 33, 415

Fraley, C. 1998, SIAM J. Sci. Computing, in press

Hartigan, J. A. 1975, Clustering Algorithms (NY:Wiley)

Horack, J. M. \& Hakkila, J. 1997, ApJ, 479, 371

Jain, A. K. \& Dubes, R. C. 1988, Algorithms for Clustering Data (Englewood
Cliffs NJ:Prentice-Hall)

Jobson, J. D. 1992, Applied Multivariate Data Analysis, 2 vols. (NY:Springer)  

Johnson, R. A. \& Wichern, D. W. 1992, Applied Multivariate Statistical
Analysis, 3rd ed., (Englewood Cliffs NJ:Prentice Hall)

Kass, R. E. \& Raftery, A. E. 1995, J. Amer. Stat. Assn., 90, 773 

Katz, J. I. \& Canel, L. M. 1996, ApJ, 471, 915

Kaufman, L. \& Rousseeuw. P. J. 1990, Finding Groups in Data (NY:Wiley)  

Kouveliotou, C., Meegan, C. A., Fishman, G. J., Bhat, N. P.,  Briggs, M. S.,  
Koshut, T. M., Paciesas, W. S. \& Pendleton, G. N. 1993, ApJ, 413, L101
(K93)

Lamb, D. Q. 1995, PASP, 107, 1152

Lamb, D. Q., Graziani, C. \& Smith, I. A. 1993, ApJ, 413, L11   

Lee, T. T. \& Petrosian, V. 1997, ApJ, 474, L37

Lestrade, J. P. 1994, ApJ, 429, L5

Mallozzi, R. S., Paciesas, W. S., Pendleton, G. N., Briggs, M. S., Preece, R.
D., Meegan, C. A. \& Fishman, G. J. 1995, ApJ, 454, 597 

MathSoft, Inc. 1996, http://www.mathsoft.com/splus

McLachlan, G. \& Basford, K. 1988, Mixture Models: Inference and
applications to clustering, New York:Marcel Dekker

Meegan, C. A., et al. 1996, ApJS, 106, 65

M\'esz\'aros, P. \& Rees, M. J. 1993, ApJ, 405, 278

Murtagh, F. 1992, in Statistical Challenges in Modern Astronomy, edited by E.
D. Feigelson \& G. J. Babu, New York:Springer, p. 449

Murtagh, F. \& Heck, A. 1987, Multivariate Data Analysis (Dordrecht:Kluwer) 
  
Murtagh, F. \& Raftery, A. E. 1984, Pattern Recognition, 17, 479  

Norris, J. P., Hertz, P., Wood, K. S. \& Kouveliotou, C. 1991, ApJ, 366, 240 

Norris, J. P., Bonnell, J. T., Nimiroff, R. J., Scargle, J. D., Kouveliotou, C., 

Paciesas, W. S., Meegan, C. A. \& Fishman, G. J. 1995, ApJ, 439, 542  

Panaitescu, A. \& M\'esz\'aros, P. 1998, ApJ, 492, 683

Pendleton, G. N., Paciesas, W. S., Briggs, M. S., Koshut, T. M., Fishman, G.
J., Meegan, C. A., Wilson, R. B., Harmon, A. B. \& Kouveliotou, C. 1994,
ApJ, 431, 416

Pendleton, G. N., Paciesas, W. S., Briggs, M. S., Preece, R. D., Mallozzi, R.
S., Meegan, C. A., Horack, J. M., Fishman, G. J., Band, D. L., Matteson, J.
L.,  Skelton, R. T., Hakkila, J., Ford, L. A., Kouveliotou, C. \& Koshut, T. M.
1997, ApJ, 489, 175

Roeder, K. \& Wasserman, L. 1997, J. Amer. Stat. Assn., 92, 894

SAS Institute Inc. 1989, SAS/STAT User's Guide, Version 6, Fourth Edition, 2
volumes, (Cary NC:SAS Institute Inc.)

Schwarz, G. 1978, Annals of Stat., 6, 461

Swayne, D. F.., Cook, D. \& Buja, A. 1991, User's Manual for XGobi,  a
Dynamic Graphics Program for Data Analysis Implemented in the X Windows
System, BellCore Tech. Memo.

Ward, J. H. 1963, J. Amer. Stat. Assn., 58, 236

White, R. L. 1997 in Statistical Challenges in Modern Astronomy II, edited by
G. J. Babu \& E. D. Feigelson, New York:Springer, p. 135

Wilks, S. S. 1932, Biometrika, 24, 471

\end{references}
\end{document}